\RequirePackage{fix-cm}
\documentclass[smallextended]{svjour3}       
\smartqed  
\usepackage{graphicx}
\usepackage{amssymb}
\usepackage{amsmath}
\usepackage{latexsym}
\usepackage{verbatim}
\usepackage{epsfig}
\usepackage{pstricks}
\usepackage{stmaryrd}
\usepackage{rotating,graphicx}
\usepackage[english]{babel}
\usepackage{setspace}
\usepackage[english]{babel}
\usepackage[utf8]{inputenc}
\usepackage[T1]{fontenc}
\usepackage{lmodern}
\journalname{Foundations of Physics}
\begin{document}

\title{Benefits of Objective Collapse Models for Cosmology and Quantum Gravity}


\author{Elias Okon \and Daniel Sudarsky}

\institute{E. Okon \at
              Instituto de Investigaciones Filosóficas, Universidad Nacional Autónoma de México\\
              Circuito Maestro Mario de la Cueva s/n, C. U., México D.F. 04510, México. \\
              \email{eokon@filosoficas.unam.mx} 
              \and
              D. Sudarsky \at
              Instituto de Ciencias Nucleares, Universidad Nacional Autónoma de México\\
               A. Postal 70-543, México D.F. 04510, México. \\
              \email{sudarsky@nucleares.unam.mx}           
}

\date{Received: date / Accepted: date}

\maketitle

\begin{abstract}
We display a number of advantages of objective collapse theories for the resolution of long-standing problems in cosmology and quantum gravity. In particular, we examine applications of objective reduction models to three important issues: the origin of the seeds of cosmic structure, the problem of time in quantum gravity and the information loss paradox; we show how reduction models contain the necessary tools to provide solutions for these issues. We wrap up with an adventurous proposal, which relates the spontaneous collapse events of objective collapse models to microscopic virtual black holes.
\end{abstract}

\section{Introduction}

An honest application of quantum mechanics to cosmology requires, by necessity, the use of an observer independent interpretation of the theory. That is, a version of the quantum formalism not fundamentally based on the notion of measurement or on that of an observer external to the studied system. The standard interpretation, then, is inadequate in this context because it relies too heavily either on measurement as a primitive term or on a division of the systems and processes of the world into macroscopic and microscopic (or observer/observed, classical/quantum, irreversible/reversible, conscious/unconscious, etc...). On the other hand, there are nowadays a number of versions of the quantum formalism available which overcome these limitations of the standard theory. This work focuses on one such group of interpretations, namely, objective collapse models, and exhibits various benefits of adopting these theories for the resolution of long-standing problems in cosmology and quantum gravity. In particular, we examine applications of objective collapse theories to three important issues: 
\begin{itemize}
\item The origin of seeds of cosmic structure
\item The problem of time in quantum gravity
\item The information loss paradox
\end{itemize}
all this with the hope of eliminating them form the current list of unresolved issues. Let's start by delineating the above-mentioned problems, along with some of the advantages of looking at them from the point of view of objective collapse models.

First we consider the quantum origin of cosmic structure and note that the standard account of such process implicitly assumes a transition from a symmetric quantum state to an essentially classical non-symmetric scenario. We claim, however, that a detailed understanding of the process that leads, in the absence of observers or measurements, from the quantum mechanical fluctuations to the classical inhomogeneities is lacking, rendering the standard account unsatisfactory (see \cite{Per.Sah.Sud:05}). The spontaneous reductions of objective collapse models, in contrast, provide explicit observer independent mechanisms for transitions from symmetric to non-symmetric states to occur. This may not only allow for a satisfactory account of the origin of cosmic structure but also may provide, through comparison with data from the Cosmic Microwave Background (CMB), with valuable clues for a better understanding of interpretational aspects of quantum theory.

Next we deal with the, so called, problem of time in quantum gravity, which emerges from the broad disparity between the way the concept of time is used in quantum theory and the role it plays in general relativity. As a result of this, at least according to an important class of theories, the ``wave function of the universe'' does not seem to depend on time, rendering time non-existent at a fundamental level. Application of objective collapse models to quantum gravity, however, may dissolve the problem by providing objective means to anchor time fundamentally.

Finally, we attack the black hole information paradox which arises from an apparent conflict between Hawking's black hole radiation and the fact that time evolution in quantum mechanics is unitary. The problem is that while the former suggests that information of a physical system falling into a black hole disappears (because, independently of the initial state, the final one will be thermal), the latter implies that information must be conserved. It is evident, however, that the paradox depends crucially on assuming a purely unitary quantum evolution. Therefore, by adopting an objective collapse model, the paradox evaporates. In what follows we will look in detail into each of the problems mentioned above and examine the advantages of evaluating them from the objective collapse perspective.

The structure of the paper is as follows: in section \ref{OC} we review the motivations and some basic features of objective collapse theories, putting special emphasis on recent fully relativistic dynamical reduction proposals. Section \ref{SCS} describes fatal flaws in the standard account for the origin of cosmic structure and delineates how objective collapse models are able to overcome them. In sections \ref{TQG} and \ref{BH} we discuss the problem of time in quantum gravity and the information loss paradox, respectively, advancing in both cases possible solutions involving objective collapse. Section \ref{Circular}  extracts methodological lessons from the analysis and outlines a possible link between spontaneous collapses and black holes. Finally, in section \ref{Sum} we round up the discussion. 
\section{Objective collapse theories}
\label{OC}
The main motivation behind objective collapse (or dynamical reduction) theories is to construct a version (or slight modification) of quantum mechanics which overcomes \emph{the measurement problem}. That is, a quantum theory which does not require for its application the notion of measurement as a fundamental and unanalyzable term. In order to achieve this, the proposal is to modify the dynamical equation of the standard theory, with the addition of stochastic and nonlinear terms, such that the resulting theory is capable, on the basis of a single law, of accounting for both the behavior of microscopic and macroscopic systems. In particular, the aim is to be able to explain, in a unified way, the quantum behavior of micro-systems and the absence of superpositions at the macro-level (without ever having to invoke observers or measurements). 

Theories with the above characteristics have been successfully constructed. The first to suggest the idea of dynamical collapse seem to have been Bohm and Bub \cite{BB}. Later, in \cite{Pea:76,Pea:79}, Pearle introduced a stochastically driven modified Schrodinger equation. Some problems with Pearle’s proposal where then solved by Ghirardi, Rimini and Weber (GRW) in \cite{GRW:85,GRW:86}. Since then, these models, along with later variations and refinements, constitute viable alternatives to the standard interpretation, (experiments that discriminate between the two are possible in principle but, for the moment, beyond technological possibilities; see \cite{Bassi} and references therein for  possible tests of collapse models and on constrains on their parameters). The main drawback with these models, however, rests in their non-relativistic nature. This is of course a substantial problem on its own but it appears even more debilitating with applications to cosmology or quantum gravity in mind. Luckily for us, recently the first fully relativistic formulations of objective collapse have been successfully constructed \cite{Tum:06,Bed:11}, allowing for the objective collapse program to be taken seriously for the applications considered in this paper.

The original GRW model \cite{GRW:86} modifies the standard quantum mechanical unitary dynamics to construct a single law which governs all physical processes, micro and macro. Furthermore, it assures that the micro-macro interactions (like those taking place in measurement processes) lead to collapses, disallowing as a consequence superpositions of macroscopic objects. In order to do so, GRW introduces spontaneous processes that occur at the microscopic level that tend to suppress superpositions of differently localized states. In more detail, it proposes that each elementary particle suffers, at random times, sudden localization processes around appropriate positions. Then, for a system of $N$ distinguishable particles with wave-function $\psi \left( \mathbf{q}_1,\mathbf{q}_2, ..., \mathbf{q}_N \right)$ the $i$-th particle may suffer a spontaneous localization around $\mathbf{x}$ in which case the wave-function gets multiplied by a (appropriately normalized) Gaussian $G_i \left( \mathbf{q_i, \mathbf{x}} \right) = K e^{-\frac{1}{2a}\left(\mathbf{q}_i-\mathbf{x} \right)} $, where $a$ is a parameter of the theory which controls the localization scale. The probability density $P(\mathbf{x})$ for the position of the center of the Gaussian is given by the square of the norm of the state $ G_i \left( \mathbf{q_i, \mathbf{x}} \right) \psi \left( \mathbf{q}_1,\mathbf{q}_2, ..., \mathbf{q}_N \right) $, which implies that there is a higher probability for localizations to occur around points where, in the standard interpretation, there is a higher probability of finding the particle. Finally, it is postulated that spontaneous localizations occur at random times distributed according to a Poisson distribution with mean frequency $\lambda$. The localization process, then, tends to destroy superpositions of well localized states with centers separated by distances greater than $a$.

In order to understand how the GRW mechanism implies the suppression of macroscopic superpositions, we note that the spontaneous localization frequency is enhanced by increasing the number of particles involved. Furthermore, when a macroscopic object is placed on a superposition of different positions, the localization of any of its constituents leads to the localization of the whole object. If follows then that macroscopic objects do not last for long superposed.

The GRW model presented above is successful in many respects. However, its original formulation is not suitable to deal with systems containing identical constituents. In order to do so one could relate the collapses not to the particles directly but to the particle number density averaged over an appropriate volume (see \cite{Tum}). Another option to deal with identical particles is to consider an alternative formulation of dynamical reduction models that goes by the name of Continuous Spontaneous Localization or CSL (see \cite{Pea:89,GPR:90}). The basic physical ideas behind CSL are the same as those of GRW; the difference is that CSL replaces the discontinuous jumps with a continuous stochastic evolution (in fact, as has been shown in \cite{GPR:90}, any CSL dynamics can be well approximated by some GRW-type one). The stochastic evolution of CSL is given by
\begin{equation}
\label{CSL}
\frac{d|\psi_w \left(t\right) \rangle}{dt}=\left[ - \frac{i}{\hbar} H + A w(t) - \lambda A^2\right] |\psi_w (t) \rangle ,
\end{equation}
with $\lambda$ a constant, $A$ a Hermitian operator (usually with position-localized states as eigenstates) and $w(t)$ a complex stochastic process. The probability for a particular stochastic process $w(t)$ to drive the evolution of an individual member of an ensemble is taken to be
\begin{equation}
P[w(t)]=\frac{1}{N}e^{-\frac{1}{2\lambda}\int_0^t w^2(\tau) d\tau} \| |\psi_w (t) \rangle\|^2 ,
\end{equation}
with $N$ a normalization factor. Therefore, if we consider a homogeneous ensemble described by the initial wave-function $|\psi (0) \rangle$, then the CSL evolution will drive individual members of the ensemble into one of the eigenvectors of $A$, each with the appropriate probability.

Even though the original CSL theory of 1989 is non-relativistic, a Lorentz-invariant generalization to quantum field theory was proposed soon after in \cite{Pea:90,RCSL:90}. It consists of a field theory for a fermion coupled to a meson scalar field supplemented with stochastic and nonlinear terms. However, such theory suffers form a renormalization problem, yielding infinite energy for the mesons, so it cannot be considered satisfactory. Recently though, important advances have occurred that suggest that theories of these type might be reconciled with special relativity. In \cite{Tum:06}, a relativistic version of GRW for non-interacting distinguishable particles was successfully constructed and in \cite{Bed:11} a model similar to the proposal in \cite{Pea:90}, but without its complications, was developed. Reference \cite{Bed:11} resolves the problems with the original proposal in \cite{Pea:90} by introducing an auxiliary field which smears out the interactions, constructing thus a fully relativistic field theory with objective collapse.

The above contributions are indeed very promising advances. However the point we want to make here is that by going beyond special relativity, and into the full general theory of relativity, collapse theories might not only be made more appealing (something that Roger Penrose has been advocating for a long time) but also that these type of theories might hold the key for the resolution of various problems and apparently paradoxical conclusions that have plagued attempts to bring together quantum theory and gravitation. In particular, as we said above, we have in mind three problems. Two of them are very well known:  the problem of time in canonical approaches to quantum gravity and the so called black hole information loss paradox. The third one is a much less noted one: the breakdown of some symmetries in cosmological situations where i) the dynamics as well as the initial conditions are symmetric, and where ii) there are no external observers, measuring devices or environments that might reasonably be called upon as providing the symmetry breaking mechanism. In the next three sections we will describe the aforementioned problems and will offer discussions of how  dynamical reduction theories might hold the key to their resolution.\footnote{Before moving on we would like to say something about the interpretation of objective collapse theories. As we mentioned before, the main motivation for the construction of these theories is to avoid the measurement problem. That is, to construct a theory without the standard probabilistic interpretation of the quantum state. However, if one removes the standard probabilistic interpretation, \emph{without} substituting it by something else, one loses the ability to make predictions and to get in touch with the physical world (see \cite{Alb.Loe:96,Lew:05}). Therefore, objective collapse models \emph{require} a new interpretation. One could think of interpreting the wave-function directly as physical, as Schr\"{o}dinger initially intended. However, the fact that wave-functions of multi-particle systems live in configuration space is something that renders this option unattractive to many. An alternative, first presented in \cite{Mass}, is to interpret the theory as describing a physical field $m(\mathbf{x},t)$, constructed as the expectation value of the mass density operator on the state characterizing the system (the relativistic version of this interpretation is discussed in \cite{MR}). Yet another option, proposed in \cite{Bel:87} and used in \cite{Tum:06}, is to take the GRW collapses, which occur at precise space-time points, as the quantities on which physical descriptions should be based.}

\section{Seeds of cosmic structure}
\label{SCS}
Thanks to some amazing technological achievements of the last decade, contemporary cosmology has not only become a precision observational discipline but also has been able to enter into regimes that were deemed beyond the reach of empirical research just a few decades ago. The inflationary era, which is thought to involve energy scales close to the grand unification regime, is now subject to a very high degree of indirect exploration and precision analysis. But not only that. As we will see below, its study unexpectedly also brings us face to face with an issue closely related to the foundations of quantum theory (a surprising development for a subject as phenomenological as this one).

The basic idea behind inflation, originally proposed to address some naturalness issues afflicting the old version of Big Bang cosmology, is that in the very early stages of its evolution the universe underwent a period of violent and accelerated expansion that erased essentially all pre-existing features. Such accelerated expansion lead to a completely homogeneous and isotropic, spatially flat ($k=0$), FRW space-time, with all matter fields, except the inflaton, in their vacuum states.\footnote{These are usually taken to be the Bunch Davies vacuum, which is a state of the quantum field naturally associated with the early de Sitter phase of the  accelerated expansion.} The inflationary era ended in a process known as ``reheating'' in which the energy stored in the potential of the inflaton field was transferred to the degrees of freedom representing ordinary matter (such as nucleons, electrons, neutrinos, photons, as well as the dark component) leading to a regime similar to that in the early stages of the traditional hot big bang cosmology.
 
It is clear, however, that without the ability to predict the emergence of structure, the inflationary paradigm would have to be dismissed; a completely featureless universe is in serious conflict not only with observations but also with our own existence (after all, such universe would lack galaxies, stars, planets and all the basic elements for the evolution of life). The great success of the inflationary paradigm lies in the fact that from such a simple and featureless stage, via the quantum fluctuations of the inflaton field, it is possible to predict the emergence of the primordial inhomogeneities and anisotropies that correspond to the seeds of all structure we see today in the universe (which is of course not homogeneous and isotropic, except in a rudimentary ``large scale average'' sense). Moreover, the predictions of the inflationary paradigm are essentially in complete agreement with the observed large scale structure of the universe as well as with the first traces of such structure imprinted in the detailed features of the CMB (beautifully mapped by the recent satellite missions of WMAP and Planck).

There is, however, an important complication with the above, apparently unblemished, success story. The problem is that the serious interpretational difficulties of quantum theory, usually associated with the measurement problem, make their appearance, and they do so in a particular disturbing form: one in which the standard strategies one normally uses to evade such questions are simply not available. Before examining this problematic aspect, though, we present a brief outline of the analysis that leads, within the inflationary paradigm, to the so called prediction of the spectrum for the primordial perturbations that seed all the universe's structure.
\subsection{The standard story and its flaws}
The theory describing the inflationary regime characterizes space-time in terms of the general theory of relativity and the matter content of the universe in terms of a scalar field called the inflaton (as already mentioned, all other fields are taken to be in an innocuous vacuum state as the result of the early stages of inflation). The theory is thus specified by the Einstein-Hilbert action for the metric and that of a scalar field with potential $V$ for the inflaton:
\begin{equation}
 S= \int d^4 x \sqrt{-g} \lbrace \frac{R}{16\pi G } + \nabla^\mu \phi \nabla_\mu \phi + V(\phi) \rbrace.
 \end{equation} 

As prescribed by the basic inflationary paradigm, one considers an homogeneous and isotropic background corresponding to a flat FRW space-time:
 \begin{equation} 
 ds^2 = a(\eta)^2 \lbrace - d\eta^2 + \delta_{ij} dx^i dx^j \rbrace,
 \end{equation} 
 where $\eta $ is known as the conformal time, and $\lbrace x^i \rbrace $ are the 3 spatial coordinates. Similarly the matter sector is described by the homogeneous and isotropic background scalar field $ \phi = \phi_0(\eta)$.
 
 The evolution of space-time is described by the relevant Einstein equation: 
 \begin{equation}
3\mathcal{H}^2=4\pi G (\dot{\phi}_0^2+2a^2 V_0),
\end{equation}
where $\mathcal{H}\equiv \dot a/ a $ represents the expansion rate of the universe and where $`` \ \dot{}  \ "\equiv \frac{\partial }{\partial \eta}$ ,
 while the scalar field satisfies the Klein Gordon equation in this background:
 
 \begin{equation}
 \ddot\phi_0(\eta) -2\dot \phi_0(\eta) \mathcal{H} + \frac{\partial V}{\partial \phi}=0.
\end{equation}
The classical background corresponding to an attractor solution of these equations is known as a ``slow roll'' (analogous with the terminal velocity regime of a gravitational fall of an object immersed in a fluid). 

Next one considers the perturbations which are described as follows: the complete scalar field is written as $\phi(x)=\phi_0(\eta)+\delta \phi(\eta,\vec{x})$ with
$\delta \phi(\eta,\vec{x})\ll \phi_0(\eta)$,
while the space-time metric is (working in the Newtonian gauge and restricting attention to the so called scalar perturbations) expressed as:
 \begin{equation}
ds^{2}=a^{2}(\eta)\left[-(1+2\psi)d\eta^{2}+(1-2\psi)\delta_{ij}dx^{i}dx^{j}\right],\quad\textrm{with}\quad\psi(\eta,\vec{x})\ll 1\label{FRW perturbed}.
\end{equation}
The perturbations $\delta \phi$ and $\psi$ are then conveniently described in terms of the new variables: 
 
\begin{equation}\label{defs}
u\equiv \frac{a\psi}{4\pi G \dot{\phi}_0},\quad v\equiv a\left(\delta\phi+\frac{\dot{\phi}_0}{\mathcal{H}}\psi\right).
\end{equation}

The next step is to give a quantum characterization of these perturbations writing the quantum field in terms of creation and annihilation operators:
\begin{equation}
 \hat{v}(x,\eta )=\sum_{\vec k} \left(\hat{a}_{\vec k} v_{\vec k} (\eta)e^{i\vec k.\vec x} +\hat{a}^{\dagger}_{\vec k} v^*_{\vec k} (\eta)e^{-i\vec k.\vec x} \right),
\end{equation}
from which one evaluates the two point quantum correlation function given by $\langle0| \hat{v}(x,\eta ) \hat{v}(y,\eta ) |0 \rangle $ to extract the so called ``Power spectrum'':
 
\begin{equation}
 \langle0| \hat{v}(x,\eta ) \hat{v}(y,\eta ) |0 \rangle = \int d^3k e^{ ik (x-y)} P(k)/k^3.
\end{equation}
 
The quantity $P(k)$ is then taken to characterize the statistical features of the primordial inhomogeneities and anisotropies in our universe. More precisely, when considering our universe as an element of an ensemble of similar universes, one takes this $P(k)$ to characterize the Fourier transforms of ensemble averages such as $\overline {\psi (x) \psi(y)}$ (the bar denotes ensemble average). Further considerations lead to an identification of such averages over ensembles of universes with suitable spatial averages, and, ultimately, with orientation averages over our own universe, \cite{Statistics}.  And, as we remarked above, when all this is done, taking into account non-trivial but well understood dynamical information about the behavior of the content of the universe from reheating to recombination, the results of the analysis lead to predictions that are in exquisite agreement with detailed observations, particularly those emerging from the satellite maps of the CMB.

Nevertheless, and in spite of the apparent success, the problem we mention above remains. It becomes evident when one questions how, from a regime that was described both at the classical and quantum level as completely homogeneous and isotropic, arose, via a dynamics that is not supposed to include anything capable of breaking said symmetries, a situation that includes the small deviations from homogeneity and isotropy that characterize the seeds of structure in the universe. One might dismiss this issue noting that we are used to a symmetric situation leading to asymmetric results in many examples involving quantum theory. For instance, in the case of a double slit experiment, when focussing attention to a single particle going trough the arrangement, we know that the spot on the screen will appear either to the right or left of the center, despite the left-right symmetry of the initial arrangement. Similarly, when considering a determination of the position of an harmonic oscillator prepared on its ground state, we all know the high likelihood of obtaining a value that differs from the only symmetric result corresponding to the origin, despite the symmetry in both dynamics and initial state of the system. However, the point we want to emphasize is that all such situations involve \emph{measurements}. Therefore, it is clear that the type of analysis described above relies implicitly either on the Copenhagen interpretation or on some other operational interpretation of quantum theory where special rules are employed whenever some measurement takes place.

If, in contrast, one wants to apply this type of strategy to cosmology, one runs into trouble. The problem is that such reliance on external observers or measuring devices is something to which we cannot appeal, particularly when considering a stage of the universe with no structure, no stars, no planets and no observers. Still, most inflationary cosmologists seem to take the view that there is no problem at all and attempt to address the basic issue invoking a variety of arguments. There is, however, no consensus among them on what is the solution. Some researches, in contrast, do acknowledge that there seems to be something unclear regarding this issue. For instance, \cite[p. 364]{Padmanabhan} states that one must work with {\it certain classical objects mimicking the quantum fluctuations}, and that this is not easy to achieve or justify. Moreover, some recent books on the topic clearly indicate that there is a problem. For example, in \cite[p. 476]{Weinberg},  we find ``the field configurations must become locked into one of an ensemble of classical configurations with ensemble averages given by quantum expectation values... It is not apparent just how this happens,'' and the manuscript \cite[p. 348]{Mukhanov} clearly acknowledges that the problem is not truly resolved by invoking one of the most popular arguments, namely that based on decoherence: ``However, decoherence is not enough to explain the breakdown of translational invariance.'' Nevertheless, and despite of these clear acknowledgements of the problem, most inflationary cosmologists seem to hold the belief that these issues have been successfully resolved within such an approach.
 
A recent collection of works, \cite{Per.Sah.Sud:05,othersus}, offers a series of critical analyses of the standard proposals. The conclusion reached is that all of the existing justifications fail to be fully satisfactory and that something beyond standard physics is required in order to provide a reasonable account for the success of the inflationary predictions regarding the emergence of the seeds of cosmic structure. The basic difficulties are appreciated by considering the values for the directly observable quantities $\alpha_{lm}$, which, according to the classical theory, are given by
\begin{equation}\label{alm2}
\alpha_{lm} = \frac{4 \pi i^l}{3} \int \frac{d^3{k}}{(2 \pi)^3} j_l (kR_D) Y_{lm}^* (\hat{k}) \Delta (k)
 \psi_{\vec{k}} (\eta_R),
\end{equation}
with $j_l (kR_D)$ the spherical Bessel function of order $l$, $\eta_R$ the conformal time of reheating, which can be associated with the end of the inflationary regime, and $R_D$ the co-moving radius of the last-scattering surface. The modifications associated with late time physics such as plasma oscillation are encoded in the transfer functions $ \Delta(k)$.

Now, since $\langle0| \psi_{\vec{k}} |0 \rangle = 0$, if we compute the expectation value of the right hand side of Eq. (\ref{alm2}), we obtain zero. However, it is clear that, for any given $l, m$, the experimentally observed value of this quantity is not zero. That is, if we apply the identification rule used in the standard approach to the one-point function, we find a large conflict between prediction and observation. Most advocates of the standard approach would reply to this criticism by arguing that $ \langle \alpha_{lm} \rangle =0 $ is not to be taken as ``the prediction of the approach'' regarding our universe and that such conclusion should only hold for an ensemble of universes. The problem, of course, is that they are unable to justify both this and the different position taken with respect to higher $n$-point functions used to derive the power spectrum. All this makes clear that in order to avoid confusion and to be able to judge the theory on its true merits, it is imperative to disentangle the various statistical aspects%
\footnote{Such statistical aspects include considerations about: i) an hypothetical ensemble of universes, ii) different space and time regions of our universe and iii) distinct orientations of our observations and characterizations of the CMB. Issues regarding the assumed connection between the quantum and statistical aspects of our characterization of the objects of interest also need to be considered.}
and to make explicit the assumptions underlying the  several kinds of  identifications that  are often  employed, as well as their expected limitations.

It seems clear, then,  that the standard approach could be considered satisfactory only if it is able to explain what exactly is wrong with the conclusion that given an initially symmetric state, the standard quantum evolution, controlled by a symmetric dynamics, cannot lead to anything but a symmetric state. Remember, however, that just as the Fourier transform of a function is a weighted average (with weights $e^{i\vec{k} \cdot \vec{x}}$), so are the Spherical Harmonic transforms of functions. Thus, $\alpha_{lm}$ is a weighted average over the last scattering surface (cosmologists often refer to average over the sky) because it is an integral over the celestial two sphere of $\frac{\delta T}{T}$ weighted over the $Y_{lm}$'s. The common argument in the literature, as we have noted, indicates that averaging over the sky justifies the identification of observations with quantum expectation values. In other words, the argument states that the relevant prediction (obtained in terms of quantum expectation values) refers to the ensemble averages, but that it can be identified with spatial averages, which, in turn, can be identified with averages over the sky. However, apparently, this should not hold for weighted averages over the sky, otherwise all the $\alpha_{lm}$'s would be $0$. The problem is that there seems to be no clear answer of why this is so because, as we already said, if we take the theoretical estimate as
\begin{equation}
\alpha_{lm}^{th} = \frac{4 \pi i^l}{3} \int \frac{d^3{k}}{(2 \pi)^3} j_l (kR_D) Y_{lm}^* (\hat{k}) \Delta (k)
\langle 0 | \psi_{\vec{k}} (\eta_R)| 0\rangle =0,
\end{equation}
 and compare it with the measured quantity $\alpha_{lm}^{obs} $, we find a large discrepancy.\footnote{In order to solve this problem within the standard account, one would need to hold that to allow the identifications described above it is necessary to invoke a further averaging over orientations, without which predictions are not reliable. The problem is that there is no justification whatsoever for this assumption.}

\subsection{Cosmic structure in light of collapse theories}
The basic idea behind our proposal is to address the problems raised above by modifying the standard quantum theory in such a way as to obtain a different prediction (hopefully compatible with observations) for the quantity considered of direct observational relevance:
\begin{equation}
\alpha_{lm}^{th} = \frac{4 \pi i^l}{3} \int \frac{d^3{k}}{(2 \pi)^3} j_l (kR_D) Y_{lm}^* (\hat{k}) \Delta (k)
\langle  \psi_{\vec{k}} (\eta_R) \rangle .
\end{equation}
In order to do so, we propose to include spontaneous collapses to the dynamics of the wave function so that the state of the quantum field need not remain at the initial vacuum state, evolving instead into some non-homogeneous and non-isotropic state. As a result, the LHS of the above expression need not vanish. The original works where this idea was first proposed involve a very simplistic version of dynamical collapse theories, \cite{Per.Sah.Sud:05,othersus}. Still, they will suffice in order to illustrate its basic features regarding the resolution to the problem at hand. 
 
The setting is that of semiclassical gravity where Einstein's equations read:
 \begin{equation}
G_{\mu\nu}=8\pi G\langle \hat{T}_{\mu\nu} \rangle 
\end{equation}
 and matter is described in terms of states of a quantum field, which, in our case, will be the scalar field of inflationary cosmology. As usual, we split the treatment into that of a classical homogeneous (``background'') part and an
in-homogeneous (``fluctuation'') component, i.e. {$g=g_0+\delta g$,
$\phi=\phi_0+\delta\phi$}. The background is taken again to be a FRW universe (with vanishing Newtonian potential), and a homogeneous scalar field {$\phi_0(\eta)$}; in a more precise treatment this field would correspond to the zero mode of the  quantum field  which  would be  quantized in its totality (see \cite{Alberto}). The main difference, with respect to the ordinary approach, will be in the spatially dependent perturbations since in this setting it is necessary to quantize the scalar field but not the metric perturbations. We will set {$a=1$} at the ``present cosmological time,'' and assume that the inflationary regime ends at a value of {$\eta=\eta_0$}, which is negative and very small in absolute terms. Then, the semiclassical Einstein's equations, at lowest order, lead to
{\begin{equation}\qquad
\nabla^2 \psi =4\pi G \dot \phi_0 \langle \delta\dot\phi \rangle = s \langle \delta\dot\phi \rangle , 
\end{equation}}
where {$s\equiv 4\pi G \dot \phi_0$}.

Consider now the quantum theory of the field {$\delta\phi$}. In this practical treatment it is convenient to work with the rescaled field variable {$y=a \delta \phi$} and its conjugate momentum {$ \pi = \delta\dot\phi /a $}. We decompose them as:
\begin{equation}
y(\eta,\vec{x})=
 \frac{1}{L^{3}}\sum_{\vec k}\ e^{i\vec{k}\cdot\vec{x}} \hat y_k
(\eta), 
\qquad \pi_y(\eta,\vec{x}) =
\frac{1}{L^{3}}\sum_{\vec k}\ e^{i\vec{k}\cdot\vec{x}} \hat \pi_k
(\eta),
 \end{equation}
where
$ \hat y_k (\eta) \equiv y_k(\eta) \hat a_k +\bar y_k(\eta)
\hat a^{+}_{-k}$ and
$\hat \pi_k (\eta) \equiv g_k(\eta) \hat a_k + \bar g_{k}(\eta)
\hat a^{\dagger}_{-k}$.
 The usual choice of modes
 $ y_k(\eta)=\frac{1}{\sqrt{2k}}\left(1 - \frac{i}{\eta
k}\right)\exp(- i k\eta)$, $g_k(\eta)=-
i\sqrt{\frac{k}{2}}\exp(- i k\eta)$ 
leads to  the Bunch Davies vacuum, i.e., the state defined by $\hat a_k{} |0\rangle=0$. At this point it is worthwhile to remind the reader that this state is translational and rotationally invariant, as can be easily checked by applying the corresponding rotation and displacement operators to it. Note also that $\langle 0|\hat y_k(\eta) |0\rangle=0$ and $ \langle 0|\hat \pi_k(\eta) |0\rangle=0$. As in the standard approach, this will be taken as the initial state for the inflaton. Nevertheless, the collapse events will modify the state and thus the expectation values of the operators {$\hat y_k (\eta)$} and {$\hat \pi_k (\eta)$}.

Next, we specify broad rules according to which collapse happens, leaving the door open for different concrete collapse mechanisms. These general rules will allow us to determine relevant features of the state {$|\Theta\rangle$} after collapse. Specifically, we will assume that the collapse mechanism is such that, after collapse, the expectation values of the field and momentum operators in each mode will be related to the uncertainties of the pre-collapse state (note that these quantities are \emph{not} zero for the vacuum because, for such state, {$\hat y_k$} and {$\hat \pi_k$} are characterized by Gaussian wave functions centered at {$0$} with spread {$\Delta { y_k}$} and  {$\Delta{{\pi_y}_k}$}, respectively). 

Then, in accordance to the generic form of the collapse mechanism described above, we will assume that at time {$\eta^c_k$} the part of the state corresponding to the mode {$\vec k$} undergoes a \emph{sudden jump} so that, immediately afterwards, the state describing the system is such that,
 \begin{equation}
 \langle{ \hat y_k
(\eta^c_k)} \rangle_\Theta= \ \ x_{k,1}
\sqrt{\Delta{\hat y_k}}, \qquad \langle{\hat \pi_k{}
(\eta^c_k)}\rangle_\Theta
= \ \ x_{k,2}\sqrt{\Delta {\hat \pi^y_k}},
\end{equation}
where {$x_{k,1} , x_{k,2} $} are (single specific values) selected randomly from within a Gaussian distribution centered at zero with spread one. Finally, using the evolution equations for the expectation values (i.e., using Ehrenfest's Theorem), we obtain {$\langle{\hat y_k{} (\eta)} \rangle$} and {$\langle{\hat \pi_k{} (\eta)} \rangle$} for the state that resulted from the collapse at later times (we are assuming here that only one collapse event happens for each mode).

The semi-classical version of the perturbed Einstein's equation that, in our case, leads to
$\nabla^2 \psi =4\pi G \dot \phi_0 \langle \delta\dot\phi \rangle$ indicates that
the Fourier components at the conformal time {$ \eta$} are given by:
 \begin{equation}
 \psi_k ( \eta) = -(s/ak^2) \langle{\hat \pi_k{}(\eta)} \rangle .
 \end{equation}
Then, prior to the collapse, the state is the Bunch Davis vacuum for which it is easy to see that {$ \langle 0|{\hat \pi_k{}(\eta)}|0 \rangle =0$}. Therefore, in that situation we have {$\psi_k ( \eta) =0$}.
 However, after the collapse has occurred, we have instead:
$\psi_k ( \eta) = -(s/ak^2) \langle\Theta |{\hat \pi_k{}(\eta)}| \Theta \rangle
\not=0$.
Then, from those quantities, we can reconstruct the Newtonian potential (for times after the collapse): 
 \begin{equation}
 \psi(\eta,\vec{x})=
 \frac{1}{L^{3}}\sum_{\vec k}\ e^{i\vec{k}\cdot\vec{x}} \psi_k (\eta)=\sum_{\vec k}\frac{s U(k)} {k^2}\sqrt{\frac{\hbar
k}{L^3}}\frac{1}{2a}
 F(\vec{k})e^{i\vec{k}\cdot\vec{x}},
\end{equation} where {$F(\vec{k})$} contains, besides the random quantities $x_{k, i}, i =1, 2$, the information about the time at which the collapse of the wave function for the mode {$\vec{k}$} occurred.

 We now focus our attention on the Newtonian potential on the surface of last scattering: {$\psi(\eta_D,\vec{x}_D)$}, where {$\eta_D$} is the conformal time at decoupling and {$\vec{x}_D$} are co-moving coordinates of points on the last scattering surface corresponding to us as observers. This quantity is identified with the temperature fluctuations on the surface of last scattering. Thus:
\begin{equation}
 \alpha_{lm}=\int \psi(\eta_D,\vec{x}_D) Y_{lm}^* d^2\Omega.
 \end{equation}
The factor {$U(k)$} is called the transfer function and represents known physics like 
 the acoustic oscillations of the plasma. Now, putting all this together, we find
 \begin{equation}
\qquad \qquad \alpha_{lm}=s\sqrt{\frac{\hbar}{L^3}}\frac{1}{2a} \sum_{\vec
k}\frac{U(k)\sqrt{k}}{k^2} F(\vec k) 4 \pi i^l j_l(|\vec k|
R_D) Y_{lm}(\hat k),\label{alm1}
\end{equation}
 where {$j_l(x)$} is the spherical Bessel function of the first kind, {$R_D \equiv ||\vec{x_D}||$}, and {$\hat k$} 
 indicates the direction of the vector {$\vec k$}. Note that in the usual approach it is impossible to produce an explicit expression for this quantity, which is different form zero. 

Thus {$\alpha_{lm}$} is the sum of complex contributions from all the modes, i.e., the equivalent to a two dimensional random walk, whose total displacement corresponds to the observational quantity. We then evaluate the most likely value of such quantity, and then take the continuum limit to obtain:
\begin{equation}
|\alpha_{lm}|^2_{M. L.} 
=\frac{s^2 \hbar}{2 \pi a^2} \int \frac {U(k)^2
C(k)}{k^4} j^2_l((|\vec k| R_D) k^3dk .
\end{equation}
The function {$C(k)$} encodes information contained in {$F(k)$}, and for each model of collapse it has a slightly different functional form. It turns out, however, that in order to get a reasonable spectrum, we have one single simple option:  the  quantity {$\eta_k^c k$} must be almost independent of {$k$}, leading to {$\eta_k^c=z/k$}.   This   gives in principle  an interesting constraint  on the   times  of collapse  for this particularly simple  model. In fact,  as  we  have in mind   that the collapse  corresponds to some sort of stochastic  process, it  seems difficult to envision how  could  the resulting collapse times  follow  such  a precise pattern. If we  want to take this  collapse  scheme  seriously we  would need to consider deviations from  the rule {$\eta_k^c=z/k$}.
Therefore, we have considered simple departures from the above pattern assuming {$\eta_k^c=A/k +B$}, and  have  confronted the  results   with  observations. A preliminary  study was  carried out in  \cite{Unanue:2008} and  a more realistic  analysis,  incorporating the well understood late time physics (acoustic oscillations, etc.), was carried out in  \cite{SLandau:2012}. Those works represent the first constrains on collapse models coming from CMB observations; they illustrate the fact  that   one  can  use  the inflationary  regime,  taken as the  originator  of   primordial   fluctuations, in order  to constrain specific versions of  collapse theories (.
 
Recently there have been various works involving the use of different versions of CSL specifically adapted to the cosmological question at hand, \cite{J-Martin,Pedro,T.P.Singh}. The results turn out to depend on the exact manner in which the CSL theory is adapted to the situation involving cosmology and quantum fields: the approach followed in \cite{J-Martin} leads to results that disagree strongly with the known features of the fluctuations and their spectrum, while the modifications incorporated in \cite{T.P.Singh} seem to resolve the most problematic aspects while still leading in principle to particular signatures that can be searched for in the observed data. In the approach taken in \cite{Pedro} it is found that it is possible to obtain acceptable spectra and that the various CSL versions generically lead to small deviations in the form of the predicted spectrum, a fact that opens the proposal to direct confrontation with observations. All this shows the interesting interplay between the quest to understand the conceptual problems that lead to this research path, and the search for the exact manner in which the modified quantum theory works in contexts where gravity is known to play a central role. It is clear that much is left to be studied in relation with these questions.
 
\section{Time in quantum gravity}
\label{TQG}
General relativity, emulating special relativity, treats time and space in a intrinsically unified way. Thus, the fundamental object of interest is the space-time, usually described in terms of a pair $(M, g_{ab})$ where $M$ is a differential manifold of dimension 4 and $g_{ab}$ a smooth pseudo-Riemannian metric with signature $ (-+++) $. Matter, if included,  is described in terms of appropriate tensor fields $A^{i}_{abc...}$ on the manifold. An essential aspect of such characterization is the diffeomorphism invariance: the fact that a particular physical situation can be equally well represented by the set $(M, g_{ab},A^{i}_{abc...} )$ as well as by the set $(M, \Phi_{*} g_{ab}, \Phi_{*}A^{i}_{abc...} )$ where $\Phi : M\to M$ is any diffeomorphism. To make things more transparent let's assume we have fixed once and for all some coordinates $x^{\mu}$ on the manifold. Thus, what one normally considers to be the coordinate transformations $x^{\mu} \to y^{\mu} = f^{\mu}(x)$, in our setting are represented by the diffeomorphism $\Phi_f : M\to M $ such that the point $p\in M$ is mapped to $\Phi_f (p)$ in the following way: if $ x_p^{\mu} (p)$ are the coordinates of $p$ then $x^{\mu} = f^{\mu} (x)$ are the coordinates of $\Phi_f (p)$. Given that in general relativity the space-time metric plays the role of the dynamical variable, it is necessary to disallow from the formalism any fixed structure restricting the form of $\Phi : M\to M $. %
For simplicity in the discussion, in the rest of this section we will restrict ourselves to general relativity in the absence of matter fields; such restriction does not imply any loss of generality.
 
\subsection{The problem}
When one attempts to apply the canonical quantization procedure to general relativity, the diffeomorphism invariance of the theory has some very important and problematic implications. In order to implement the quantization process, one starts with the Hamiltonian formulation of the classical theory. Such formulation corresponds to writing the theory in terms of suitable variables associated with a given spatial hypersurface $\Sigma_0$ (that corresponds to an embedded manifold in $M$), along with with a foliation $\Sigma_t$ of the space-time characterizing the evolution of such variables. More specifically, the canonical data are taken to be the 3 metric of the hypersurface $\Sigma_t$, $h_{ab}$, and its canonical conjugate momentum, $\pi^{ab}$, which is related in a simple way to the extrinsic curvature $K_{ab}$ characterizing the embedding of $\Sigma$ in $M$. The foliation is then specified in terms of the so called lapse function $N$ and shift vectors $N^a$ that determine the points of the $\Sigma_t$ and $\Sigma_{ t +\Delta t}$ hypersurfaces that are to be used in characterizing the evolution of the canonical variables. 
 
An important feature of this formulation is that, just as the initial data cannot be freely and arbitrarily specified, because it must satisfy the constraint equations, the canonical data is also subject to those constraints. These take the form:
 \begin{equation} 
 {\cal H} (h_{ab}, \pi^{ab}) =0
 \end{equation}
 and 
 \begin{equation} 
 {\cal H}_a (h_{ab}, \pi^{ab}) =0
 \end{equation}
where the terms ${\cal H}$ and ${\cal H}_a $ are specific functions of the canonical variables. These are known as the Hamiltonian and diffeomorphism constraints, respectively. The phase space of the theory, $\Gamma$, is then the set of pairs of tensor fields $(h_{ab}, \pi^{ab})$, defined on the manifold $\Sigma$, that satisfy the above constraints. 
 
 
The Hamiltonian that generates the evolution along the vector field $t^a= n^a N + N^a$ is then expressed as 
 \begin{equation} \label{Hamiltonian}
 H =\int d^3 x \sqrt h [ N {\cal H} + N^a {\cal H}_a ],
 \end{equation}
resulting in the equations of motion:
 \begin{equation} \label{Evolution equations}
 \dot{h}_{ab} = \frac{\delta H}{\delta \pi^{ab}}, \qquad \dot{\pi}^{ab}= \frac{\delta H}{\delta h_{ab}},
 \end{equation}
which are equivalent to Einstein's field equations.
Therefore, the classical equations of motion allow us to identify canonical data on any hypersurface with a full space-time metric. Note however that such identification requires an arbitrary choice of the lapse and shift functions that appear in Eq. (\ref {Hamiltonian}) because those functions must be used in the reconstruction of the space-time metric according to:
 \begin{equation} \label{reconstructed metric}
 g_{ab} = - (N^2 -N^i N^j h_{ij})(dt)_a (dt)_b - 2 h_{ij}N^j (d x^i)_a (dt )_b+ h_{ij}(t) (d x^i)_a (d x^j)_b 
 \end{equation}
 where we have taken the coordinates $\lbrace x^i \rbrace$ associated with $\Sigma$ to express the spatial metric as 
$ h_{ab} (t)= h_{ij} (t) (d x^i)_a (d x^j)_b$, and used the time parameter appearing in the evolution dictated by Eqs. (\ref{Evolution equations}) as the fourth coordinate of our reconstructed space-time. The arbitrariness in the choice of lapse function and shift vector disappears from the construction when we note that the resulting space-time metric, Eq. (\ref{reconstructed metric}), is independent of such choice (up to diffeomorphisms).
 
The situation in the canonical quantum theory is quite different. The canonical quantization procedure involves replacing the phase-space variables $h_{ab}, \pi^{ab}$ by operators in a Hilbert space $\hat h_{ab}, \hat \pi^{ab}$ , such that the basic Poisson brackets are replaced by the commutation relations via the rule $\lbrace , \rbrace \to [ , ]$. The Hilbert space is usually taken to be the space of wave functionals on the configuration variable $\Psi ( h_{ab})$ (or some equivalent alternative such as the momentum variable) with an appropriately chosen inner product (such that the operators are hermitian). At this point what we have is the unphysical  or auxiliary Hilbert space ${\mathcal H}_{Aux}$. The physical Hilbert space ${\mathcal H}_{Phys}$ is (according to the Dirac procedure for the quantization of a constrained system), the subset of ${\mathcal H}_{Aux}$ satisfying the operational constrains:
 \begin{equation} \label{quantum constrains}
 \hat {\cal H }\Psi ( h_{ab}) =0 \qquad
 \hat {\cal H}_a \Psi ( h_{ab}) =0.
 \end{equation}
 Time evolution, as usual, is controlled by the Schr\"{o}dinger equation:
 \begin{equation} \label{Schroedinger}
 i \frac{ d}{dt }\Psi ( h_{ab}) = \hat H \Psi ( h_{ab}).
 \end{equation}
However, since $ \hat H =\int d^3 x \hat {\sqrt{h}} [ N \hat {\cal H} + N^a \hat {\cal H}_a ]=0$, it is clear that the quantum constraints in Eq. (\ref{quantum constrains}) lead to a wave functional that is independent of $t$.
 
Thus, we end up in the quantum theory with wave functionals that depend only on the spatial metric $h_{ab}$ but not on ``time.'' Time has disappeared from the central objects characterizing any specific situation according to the formalism. The question that arises is the following: is there a quantum procedure that corresponds to the space-time reconstruction carried out in the classical theory? That is, is there anything in the quantum theory that corresponds to the classical procedure that leads to Eq. (\ref{reconstructed metric})? Moreover, it seems essential for the consistency with the basic conceptual underpinning of general relativity that the reconstruction so obtained be independent from whatever ends up corresponding to the slicing of space-time or the choice of lapse and shift. In other words, at the conceptual level, it seems essential that the theory and reconstruction procedure ensure the diffeomorphism invariance of the resulting characterization of whatever is taken as representing the quantum version of space-time.
 
Although we have focused the discussion above on the case of a pure gravity theory, the inclusion of matter fields does not alter the picture in any essential manner. For a theory including matter fields that do not break the diffeomorphism invariance of the full theory (such as the fields appearing in the standard model of particle physics along with other fields that might play relevant roles in our current understanding of nature such as the inflaton, or the hypothetical fields employed in most theories about dark energy or dark matter) the constraint equations take a similar form:
\begin{equation} 
 {\cal H} ={\cal H}^{g} (h_{ab}, \pi^{ab}) + {\cal H}^{m} (h_{ab}, \phi, p) =0
 \end{equation}
 and 
 \begin{equation} 
 {\cal H}_a={\cal H}^{g}_a (h_{ab}, \pi^{ab}) + {\cal H}_a^{m} (h_{ab}, \phi, p)=0
 \end{equation}
where the superscript $g$ labels the standard terms appearing in the pure gravity theory and where the new contributions ${\cal H}^{m} , {\cal H}_a^{m} $ represent terms coming from the matter sector of the theory that depend on the metric and matter variables described here generically in terms of configuration and momentum field variables $\phi$ and  $p$. The point is that the evolution Hamiltonian is, once again, given by the expression in Eq. (\ref{Hamiltonian}) so that the Schr\"{o}dinger equation will again lead to a time independent wave functional for any element of the physical Hilbert space corresponding to those elements of the auxiliary Hilbert space annihilated by the constraints. 
 
There are in the literature a large number of proposals to address this problem. However there seems to be none that can be regarded as completely satisfactory. The most popular approaches are based on the identification of some variable of the gravity-matter theory to act as a physical clock and on interpreting the wave functional in a relational manner. That is, on using the wave functional to evaluate the probability that certain observable takes a given value when the time described by the clock variable takes some other value. The scope and shortcomings of this and other approaches are discussed, for instance, in \cite{Isham}. In this context it is perhaps worth mentioning the recent observation made in \cite{Gambini-Pullin1,Gambini-Pullin2} that when one uses such physical clock variables and considers the evolution of the wave functional in the time measured by it, the resulting evolution equation resembles the standard Schr\"{o}dinger equation, but includes terms that break the unitarity of the evolution. We will not discuss those issues any further here as we want to focus on the plausibility of a very different kind of solution to the problem. 
 
\subsection{A solution involving collapse}
 As we have seen, the problem of time in quantum gravity is intimately tied with the fact that, in general relativity, characterizations of space-time (along with fields living in it), using different Cauchy hypersurfaces, are equivalent. At the classical level, this corresponds to the fact that a space-time (including all fields in it) is completely determined by Cauchy data on any given hypersurface. At the quantum level, this equivalence implies that the state characterized in terms of the appropriate variables $h_{ab} , \phi $ should be independent of the hypersurface. The problem of time is thus connected to the fact that information does not change from hypersurface to hypersurface. This situation clearly changes when a dynamical stochastic collapse is incorporated into the picture. In that case, the characterization of the situation is continuously being modified due to the random collapses. In other words, the information difference between two hypersurfaces corresponds to the collapse events taking place between the two.
 
One way to think about this is by examining the interaction between the stochastic field of the CSL model and the rest of the physical fields. One might imagine describing such situation with something akin to the interaction picture where the standard aspects of the dynamics are treated {\it a la} Heisenberg, incorporated in the evolution of the operators, while the effect of the stochastic collapse is thread as an interaction, affecting the evolution of the wave functionals. In fact, since the collapse dynamics is fundamentally stochastic, it does not seem sensible to even attempt to incorporate its effects into a Heisenberg picture: that would result in the basic operators of the theory not even being well defined.%
\footnote{In the sense that only after the stochastic variables have been converted into quantities with definite numerical values would those operators become explicitly defined.} The point is that the full evolution equation of the wave functional $\Psi ( h_{ab})$, or more likely, that of the full geometry and matter theory, would not be controlled exclusively by a Hamiltonian constructed out of the theory's constraints.
 
Therefore, Eq. (\ref{Schroedinger})  would have to be replaced by something like:
 \begin{equation} \label{Schroedinger2}
 i  d\Psi ( h_{ab}) = \left\lbrace \int dt \int d^3 x \hat {\sqrt{h}} [ N \hat {\cal H} + N^a \hat {\cal H}_a ]  + \int d^4 x \hat {\cal C}(x) \right\rbrace \Psi ( h_{ab}) ,
 \end{equation}
where $ \hat {\cal C}(x) $ is a densitized operator characterizing the effects of the collapse dynamics. That operator will generically  include  some  random functions of space-time (usually some  real  valued  random  fields as in \cite{Bed:11}) which  would  play a role not  dissimilar  from that  of  an   external  source. Under such circumstances, and even after taking into account that the constraints annihilate the physical states, there would be a non-trivial evolution of the wave functional. 
 
In fact, the fully Lorentz-invariant version of CSL developed in \cite{Bed:11} is presented precisely in an interaction-like picture, using a Tomonaga formalism. The state is attached to a Cauchy hypersurface $\Sigma$ and the evolution from one such hypersurface to another hypersurface $\Sigma'$ goes according to: 
 \begin{equation} \label{Tomonaga}
 d_x \Psi (\phi; \Sigma') = \left\lbrace  -i J(x) A(x)  d\omega_x  - (1/2) \lambda^2 N^2(x)d\omega_x 
 +\lambda  N(x)  d W_x
 \right\rbrace\Psi(\phi; \Sigma) ,
 \end{equation}
where $ d\omega_x $ is the infinitesimal space-time volume separating the hypersurface $\Sigma$ and $\Sigma'$, $\lambda $ is the CSL coupling constant (a new constant in nature), $ J(x)$ is an operator constructed out of the matter fields in question, taken to represent a matter density associated with the degree of freedom in question (for a Dirac field one writes $J(x) = \bar \psi (x) \psi(x) $), $W_x$ is a Brownian motion field, $ A(x) $ and $N(x)$ are operators that modify the state of an auxiliary quantum field, called the pointer field, which has no independent dynamics (and thus is represented by a field with operators which commute at all space-times points) and just keeps track of the past stochastic evolution of the matter fields. For more details we remit the reader to reference \cite{Bed:11} as we do not want at this point to limit our discussion to any specific version of a collapse model (despite the fact that we find CSL, and in particular this latest version, to be very promising).
  
The point we want to emphasize here is that, independently of the specific collapse mechanism, dynamical collapse theories remove the basic feature of standard quantum theory with deterministic and unitary evolution, which is at the core of the problem of time: the fact that in that approach it is difficult to make sense of the idea that ``things happen''  because the state of the quantum field on any hypersurface contains exactly the same information as the state in any other hypersurface. This feature, when combined with the diffeomorphism invariance that lies at the core of general relativity and which forces us to consider the use of different hypersurfaces to characterize physical situation simply as a choose of gauge, leads inescapably to the problem of time we have described above. Once this feature is removed ``things start to happen'' and the difference between different hypersurfaces becomes much more that a question of gauge: if a certain hypersurface $\Sigma'$ happens to lie completely to the future of a given hypersurface $\Sigma$, such fact will be characterized at the state-level by the effects of the (infinite and continuous) spontaneous collapses corresponding to the behavior of the particular realization of the stochastic field $W_x$ in the whole space-time volume separating $\Sigma$ from $\Sigma'$. This feature can be characterized in a heuristic language as corresponding to the things that happen between the two hypersurfaces. And, naturally, if we can say that things happen, we are in a scenario where there is time.
 
It is clear that these ideas need to be worked out in much more detail, a task that quite likely requires focus on a specific, sufficiently realistic, collapse theory. However, we hope to have shown that the basic elements to successfully address the problem of time in canonical quantum gravity seem to be generically present in theories involving dynamical spontaneous reduction of the quantum state.

  
\section{Black holes and information}
\label{BH}
Black holes appear to lose information. However, such conclusion seems to go against the ``common-knowledge'' that quantum evolution preserves information regarding the quantum state of a system. This, in short, is the information loss paradox. In order to understand how objective collapse models could resolve the situation, we will first review the details that lead to this apparent puzzle.
\subsection{The paradox}
Let's start by saying something about the classical theory of black holes. As we all know, gravity is always attractive; therefore, it will tend to draw matter together to form clusters. If the mass of a cluster is big enough, nothing will be able to stop the contraction until, eventually, the gravitational field at the surface of the body will be so strong that not even light will be able to escape. At such point, a region of space-time from which nothing is able to escape forms, and such region is defined to be a black hole. Its boundary, formed by the light rays that just fail to escape, is called the event horizon and, according to classical general relativity, the area of the event horizon of a black hole never decreases. Furthermore, the no-hair theorem shows that all black hole solutions (of the Einstein-Maxwell equations) are characterized by only three parameters: mass, charge and angular momentum.

The no-hair theorem suggests that when a body collapses to form a black hole, a large amount of information is lost. That is because, while the collapsing body is described by lots of parameters (type of matter, multipole moments of the initial mass distribution, etc.) the black hole is described only by a handful of numbers. At the classical level, this loss of information does not seem to be problematic because one could hold that the ``lost'' information is still inside the black hole (we will see below that this reasoning is dubious). The quantum situation, on the other hand, is very different because, as was shown by Stephen Hawking, quantum mechanics causes black holes to radiate and lose mass, apparently until they completely disappear \cite{Haw:75}. This so called Hawking radiation is derived by doing quantum field theory on the curved space-time of a collapsing black hole. In particular, one considers a scalar field on such background and, given that the space-time is time-dependent, one finds that positive solutions at past infinity turn partly negative at future infinity. Moreover, one finds a steady particle emission rate corresponding to a thermal emission at a temperature $\frac{\kappa}{2\pi}$ where $\kappa$ is the surface gravity of the black hole.

If correct, the above implies that, after all, black holes do swallow information. Independently of the characteristics 
of the material that formed the hole, the final state is thermal, and hence, virtually information-free. There 
is a complication however, since this information loss seems to imply a violation of a cherished principle of quantum mechanics, namely, \emph{unitarity}. 
Therefore, if one holds unitarity to be immutable, then the fact that black holes lose information results in a paradox.

The standard Penrose (i.e., conformally compactified, with null lines at 45º) diagram for a classical collapsing spherical body is depicted in Figure 1. 
\begin{figure}[h]
\centering
 \begin{pspicture}(4.5,5)
 \psline(1,0)(1,4)(2.25,4)(3.5,2.5)(1,0)
\pscustom[linewidth=.5pt,fillstyle=solid,fillcolor=gray]{
\pscurve(1,0)(1.4,2)(1.3,4)
\psline[liftpen=1](1,4)(1,0)}
\psline[linewidth=2pt](1,4)(2.25,4)
\psline[linewidth=.5pt](1,2.75)(2.25,4)
\rput(2.6,1){$\mathcal{I}^-$}
\rput(3.25,3.5){$\mathcal{I}^+$}
\rput(0,3.5){\small Horizon}
\psline{->}(.7,3.47)(1.65,3.47)
\rput(1.6,4.5){\small Singularity}
\psline{->}(1.6,4.35)(1.6,4.05)
\rput(-1,2){\small Collapsing body}
\psline{->}(.42,2)(1.2,2)
 \end{pspicture}
\caption{Penrose diagram for a collapsing spherical body.}
\end{figure}
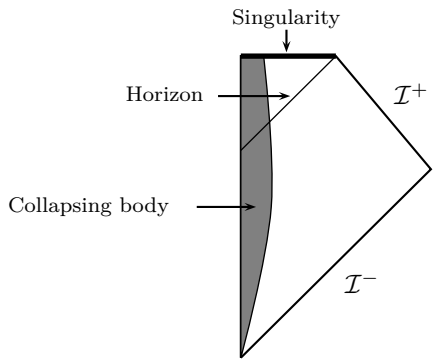
Since there are points under the horizon that are not in the past of the future null infinity $\mathcal{I}^+$, it is impossible to reconstruct the 
whole space-time by evolving backwards the data from it. In this sense, black hole formation leads to information loss even classically, and independently of the no-hair theorem. Now, in order to account for Hawking's radiation, the space-time diagram must be modified as shown in Figure 2.
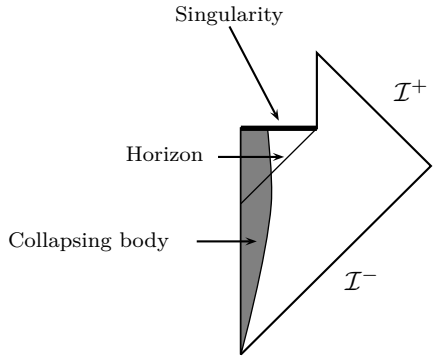
\begin{figure}[h]
\centering
 \begin{pspicture}(4.5,5)
 \psline(1,0)(1,3)(2,3)(2,4)(3.5,2.5)(1,0)
\pscustom[linewidth=.5pt,fillstyle=solid,fillcolor=gray]{
\pscurve(1,0)(1.4,2)(1.35,3)
\psline[liftpen=1](1,3)(1,0)}
\psline[linewidth=2pt](1,3)(2,3)
\psline[linewidth=.5pt](1,2)(2,3)
\rput(2.6,1){$\mathcal{I}^-$}
\rput(3.25,3.5){$\mathcal{I}^+$}
\rput(0,2.66){\small Horizon}
\psline{->}(.7,2.62)(1.57,2.62)
\rput(.85,4.5){\small Singularity}
\psline{->}(.85,4.35)(1.5,3.1)
\rput(-1,1.5){\small Collapsing body}
\psline{->}(.42,1.5)(1.21,1.5)
 \end{pspicture}
\caption{Penrose diagram for a collapsing spherical body taking into account Hawking's radiation.}
\end{figure}
In this case, there is no black hole at the end, but still, pure states at $\mathcal{I}^-$ evolve into mixed states at $\mathcal{I}^+$, signalling information loss.

In order to avoid the paradox without negating unitarity, at least three ways out have been proposed. The first consists in maintaining that Hawking's radiation does not lead to the complete disappearance of a black hole. That, at the end, a remnant containing all the information survives. The second alternative is to hold that, in the final explosion, all the information comes back out again. The last option is to claim that all the information escapes with the Hawking radiation. This last alternative has been the preferred one among particle physicists since Maldacena's work on the AdS-CFT correspondence, which suggests that a 3-d black hole evaporation process is equivalent to a 2-d quantum system without gravity. The claim is, then, that since the dual system respects the rules of ordinary quantum mechanics, information cannot be lost. 

Recently, however, a new element to the information loss paradox has been added: the ``firewalls.'' In \cite{Fire:12},\footnote{See \cite{Br1,Br2} for a similar prediction from different assumptions.} is it argued that three fundamental principles of physics cannot all be true. These principles are i) unitarity, or the fact that the black hole information is carried out by the Hawking radiation; ii) effective field theory, or the fact that physics works as expected far away from a black hole even if it breaks down at some point within it; and iii) the equivalence principle which holds that an observer falling into a black hole sees nothing special at the horizon. Furthermore, \cite{Fire:12} claims that if unitarity is to be preserved, then the equivalence principle must break down at the horizon. The idea is that, on the one hand, for Hawking's radiation to occur, the emitted particles must get entangled with the twins that fall into black hole. On the other hand, if information is to come out with the radiation, then each emitted particle must also get entangled with all the radiation emitted before it. Yet, the so called ``monogamy of entanglement'' holds that a quantum system cannot be fully entangled with two independent systems at the same time. Therefore, in order to preserve the equivalence principle, each particle needs to be entangled with its in-falling twin, but in order to preserve unitarity, the emitted radiation must be entangled with radiation that escaped at earlier times, and both cannot happen simultaneously. In order to resolve the conflict, \cite{Fire:12} decides to maintain unitarity and to break the link between escaping and in-falling pairs. The cost for this, however, is the release of an enormous amount of energy, turning the event horizon into a firewall that burns anything falling through.

A very different attempt at a dissolution of the information loss paradox comes from quantum gravity. In particular, from the hope that quantum gravity effects can resolve the singularity of the black hole, thus restoring unitarity. In \cite{Ash.Boj:05}, for example, it is claimed that given that one expects quantum gravity to be free of infinities and that the event horizon is a highly global entity, one should not expect Figure 2 to be a reliable representation of the whole physical process, and to be {\it completely unreliable} at the end point of the evaporation. The suggestion then is to substitute Figure 1 with a ``quantum space-time diagram,'' depicted in Figure 3 (see \cite{Ash.Boj:05} for details), which extends the geometry \emph{beyond the singularity}. Since the new space-time has no future boundary at the singularity, unitarity is, allegedly, restored.
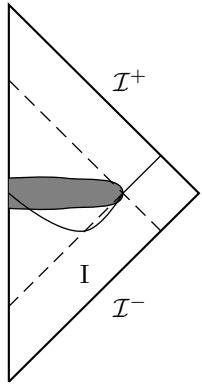
\begin{figure}[h]
\centering
 \begin{pspicture}(4.5,5)
 \psline(1,0)(1,5)(3.5,2.5)(1,0)
\pscustom[linewidth=.5pt,fillstyle=solid,fillcolor=gray]{
\pscurve(1,2.3)(1.75,2.3)(2.5,2.5)(1.75,2.7)(1,2.7)}
\psline[linewidth=.5pt,linestyle=dashed](1,1)(2.5,2.5)
\psline[linewidth=.5pt](2.5,2.5)(3,3)
\psline[linewidth=.5pt,linestyle=dashed](1,4)(3,2)
\rput(2.6,1){$\mathcal{I}^-$}
\rput(2.6,4){$\mathcal{I}^+$}
\rput(2,1.43){I}
\pscurve[linewidth=.5pt](1,2.5)(2,2)(2.5,2.5)
 \end{pspicture}
\caption{``Quantum space-time diagram'' for a black hole.}
\end{figure}
Figure 3, then is supposed to reflect the fact that quantum gravity effects have resolved the classical singularity and so the gray region represents the ``deep Planck regime'' where space-time is genuinely quantum. Again, since in this scenario there is no singularity, information is never lost and unitarity is preserved during the whole process.

The picture that emerges is then one where the early part of the evaporation process is well described by the semiclassical treatment of Hawking (perhaps with small modifications, taking into account the back-reaction in the regime where quantum gravity effects might be ignored) and which leads in region I to a seemingly non-unitary evolution. However, according to the picture, full unitarity will be restored when taking into account the radiation emitted at late times, which would have to be strongly entangled with the radiation reaching ${\mathcal{I} }^+$ at early times. The problem of course is that it is very difficult to envision exactly how such entanglement would occur, without profoundly disturbing the region of space-time where the semi-classical treatment is thought to be reliable. As we mentioned above, the recent analysis in \cite{Fire:12} indicates that the existence of an entanglement between early and late stages of the radiation would require a divergent energy momentum tensor, creating at the, would be, event horizon a ``firewall.''  As far as we know, this issue has not been completely settled but we believe it exemplifies the difficulties one must face in attempting to produce a consistent picture where black hole evaporation leaves no quantum gravity remnant and where information is not lost with unitarity preserved.%
\footnote{There is a view according to which one need not worry about information loss and unitarity breakdown simply because ${\cal I }^+$, or space-like hypersurfaces approaching it, are no Cauchy hypersurfaces. If that is the case, the singularity (or a space-like hypersurfaces near it) represents an additional part of the boundary of space-time where the causal curves can end. This is, from the purely mathematical point of view, complete accurate and correct. However, the issue is how are the observers that witness the complete evaporation, of say, a very small black hole, to characterize what they see as their space-time in which the black hole has disappeared?  What would they say regarding the evolution of the physical objects that for them represent the totality of what exist at a certain region they consider to be a Cauchy hypersurface? That means that at the effective level, that of the laws of physics as they observe them to hold, they must confront a problem: are the laws of evolution compatible with quantum unitarity or are they not?}
 
What we propose is to consider the possibility that the laws of evolution are not compatible with quantum unitarity. That information is lost and that unitary evolution is broken, but not as an extraordinary situation involving the exotic process of black hole evaporation, but that this takes place, albeit in a smaller degree, in all situations and at all times. That is, we want to propose a view in which the standard quantum theory must be modified, incorporating at the fundamental level a degree of randomness and a departure from unitarity. In any case, below we will show how objective collapse theories have no problem in accepting the fact that black holes really lose information.

\subsection{Objective collapse and the loss of unitarity}
From the above discussion it should be clear that the assessment regarding the evaporation of a black hole, including the loss of information, changes dramatically once one adopts the view that at the fundamental level neither is information conserved nor is the fundamental evolution of quantum systems describable in terms of a unitary transformation. Furthermore, the adoption of such theory allows for the black hole  evaporation process to be viewed as a particular consequence of the fundamental evolution laws of nature that now include essential non-unitarity and irreversibility. 
 
The interesting question of course is the following: is it possible to produce a theory where the information paradox is resolved not only at the qualitative level considered above but in a more precise quantitative characterization? And, more particularly, is it possible to explain the loss of information and the departure from unitary (to the extent that they are required if the standard black hole evaporation via thermal emission is maintained) in terms of a fundamental quantum evolution that includes spontaneous reductions of the wave function?

At this point, we cannot really answer these questions as that would involve not only the full development of a specific theory but its application to the case of a realistic collapsing and evaporating black hole (something that is evidently well beyond the scope of this paper and of the present status of our theoretical tools). In other words, to answer these questions in the positive would only be possible if one already has a candidate for the full theory at hand and is able to perform the desired calculation using it. What we will describe below, then, should be seen as a very schematic characterization of the theory that we envision as being able to succeed. Of course, some of its features will have to be adjusted in the course of the analysis of the problem at hand: the formation and subsequent evaporation of a black hole.
 
What we will do then is to first specify, in a schematic way, the essential characteristics of a collapse theory that might be able to model the situation. Next, we will describe a method by which one might constrain some specific features of the proposed theory by applying it to the formation and evaporation of a black hole (taking into account the specific degree of information loss and deviation from unitarity evolution). As is natural for the relativistic context at hand, we will only consider collapse theories formulated in ways that make them applicable to the realm of quantum fields, e.g., \cite{Bed:11}.\footnote{Quantum field theory is usually considered the essential language in which to handle quantum matter in a spacial relativistic context. In the regime where gravity becomes important, we also envision a quantum field theoretic treatment of matter fields but at the very fundamental level we envision some sort of quantum gravity theory, perhaps resembling some of the currently popular programs, but modified to include effects that at the more effective level look like dynamical reductions.} 

We want to consider now the treatment of the black hole formation and evaporation processes from the point of view of collapse models. Given that a correct characterization of the gravitational back-reaction, leading to the shrinking of the black hole, presumably involves a quantum treatment of the space-time degrees of freedom, i.e., a quantum theory of gravity, we will not attempt to characterize it here. Moreover, and as is often done in the literature, we will consider such issues to be, to a large extent, independent from the questions that do concern us here (i.e., information loss and lack of unitarity). We will therefore assume that as a result of including the back-reaction to the Hawing evaporation, in the effective theory that emerges from quantum gravity, we end up with a space-time diagram of the type of Figure 2. We want however to follow the lead of \cite{Ash.Boj:05} and maintain that quantum gravity effects resolve the singularity. Otherwise, as in the classical case, there will be points under the horizon that are not in the past of the future null infinity $\mathcal{I}^+$, so it will be impossible to reconstruct the whole space-time by evolving the data backwards from it. The question then is if we can characterize the processes occurring in Figure 2 in terms of Figure 3. 

In order to achieve the above, the key element to consider is a modification of the quantum dynamics such that it is capable of accounting for the evolution of the quantum state of the matter field from $ {\mathcal{I} }^-$ to ${\mathcal{I} }^+$. Of course, the major obstacles in requiring that this be compatible with the evolution depicted in Figure 2 are precisely the lack of unitarity of the transformation connecting the states in $ {\mathcal{I} }^-$ and ${\mathcal{I} }^+$ and the loss of information (usually ascribed to the singularity region of Fig 2). Nevertheless, in the context of Fig 3, which, as we have argued, must be the relevant one as far as observers in our universe are concerned, the lack of unitary evolution from $ {\mathcal{I} }^-$ to ${\mathcal{I} }^+$, and the loss of information, have to be attributed to aspects of the evolution occurring in the interior of the space-time of Fig 3. In order to achieve this, we envision that most of this unitarity and information destruction must be traced to events taking place in the region occupied by the gravitationally collapsing matter in Fig 2 and the region depicted as replacing the singularity in Fig 3. For that to work we need to ensure that in those regions the effects of the CSL terms in the evolution equation result in the enormous amount of unitarity destruction and  entropy creation that characterizes the difference between the state of matter fields before and after the black hole creation and annihilation. Specifically, we should be able to obtain the evolution from the states of the matter field from $ {\mathcal{I} }^-$ to ${\mathcal{I} }^+$.

What we are demanding in order to obtain the right evolution for the matter fields requires a (hopefully slight) modification of the standard CSL model. In particular, we need to ensure that its effects becomes larger in the present context that in ordinary situations. This can be achieved by replacing the parameter $\lambda$ (see Eqs. (\ref{CSL}) and (\ref{Tomonaga})) by some function of the gravitational environment. For instance, $\lambda$ might be replaced by a function of some geometrical scalar such as the Weyl scalar given by $ W^2 = W_{abcd} W^{abcd} $ (where $W_{abcd} $ is the Weyl tensor and the indices are lowered and raised using the space-time metric). In fact it is worth noting that phenomenological analyses indicate that in order to resolve the measurement problem (the objective for which the CSL theory was formulated in the first place), the value of the parameter $\lambda$ occurring in the original proposal should depend on particle's masses, so that, for example, the effect of the CSL theory on a proton is about a thousand times bigger than its effects on an electron, \cite{Pea.Squ:94}. It seems then that something that at an effective level looks like a coupling that depends on the space-time curvature, such as the one considered here, is a very natural way to implement the kind of mass dependence required by the original CSL theory (which must be viewed as the limiting case of the kind of theory we are envisioning here).
 
The precise nature of the coupling we are proposing must be such that it ensures a match between the generation of entropy due to the dynamical collapse process and the standard estimates of entropy generation in black hole formation and evaporation. This requirement is rather non-trivial and it is far from clear that something like this can be achieved by the simple replacement of $\lambda$ by a suitably chosen function $\lambda (W^2)$, (or some other geometrical scalar). If no such function can be found, some more radical modification of the CSL theory will be required. Nevertheless, a sensible reason for choosing  the Weyl scalar as argument of $\lambda (\,\,)$, rather than, say, the  Kretschmann scalar, is that the Weyl scalar  seems to have the features that might naturally lead to an association of low entropy with the early state of the universe and a large entropy with its late-time state. That is, we find it natural to tie some of the ideas presented here with the arguments of Roger Penrose regarding the origin of the second law of thermodynamics in its generalized form (see for instance arguments regarding the {\it ``Weyl curvature hypothesis''}  in \cite{Pen:04}). Of course the rest of our argument is independent of whether such connection exists or not. The interesting point however is that if we hold that a dynamical collapse theory could resolve the information loss paradox in a complete quantitative fashion (something that seems natural when we note that such theories contain the essential features to resolve the issue at least at the qualitative level) then we would have a powerful tool to guide us in the construction of candidate theories and, if more than one is postulated, in the dispute among alternatives.

Before concluding this section we would like to compare our ideas regarding black hole information loss with a series of speculative ideas proposed by  Penrose \cite{Pen:04} and Hawking \cite{Haw:00}. Penrose considers a thought experiment consisting of a vast box containing a black hole. He then argues that the information lost into the black hole causes trajectories in phase space to converge and volumes to shrink. That is because different inputs give rise to the same output. He holds, however, that this loss of phase space volume is balanced by the quantum spontaneous collapse process since, in the quantum case, several outputs may follow from the same input. He clarifies however that this balance is to be achieved only overall since in his proposal black holes need not be present for quantum state reductions to occur.

Hawking, in turn, observes that if information is lost in macroscopic black holes, then it should also be lost in microscopic, virtual black holes. This, he thinks, implies that quantum evolution cannot be unitary since pure quantum states will inevitably become mixed. He concludes, then, that gravity induces a new level of unpredictability into physics, additional to the standard quantum uncertainty. Of course, as is well known, Hawking eventually changed his mind, maintaining that information, after all, is not lost in the black hole. He did so after performing new calculations suggesting that information can gradually get out of the black hole through the radiation.
\section{Self-consistent bootstrapping and theory construction}
\label{Circular}
When working on fundamental physics, we are used to theories being constructed in a single direction only: we specify a few fundamental laws and we use them to describe, and to derive laws that govern, more complex situations. This basic scheme underlies fields such as statistical mechanics and classical field theories such as general relativity. In the first case, the fundamental laws, which ultimately rely on the standard model of particle physics, describe the behavior of particles. Then, from those laws, statistical ones that only apply to ensembles of large number of such individual constituents are constructed. In the second case, the basic dynamics is provided by Einstein's equation, from which one can deduce rules governing complex and extended objects, such as galaxies, planetary systems, or even more impressive results such as the basic laws of black hole dynamics (also known as the laws of black hole thermodynamics, after the incorporation of results involving other considerations such as the Hawking radiation effect). We would like to emphasize that above we are referring to the existence of clearly differentiated levels in the \emph{laws} of physics and not on levels regarding constituents of matter. That is, we are pointing out that in standard practice in physics some laws are taken as basic and others are derived from them for specific situations; we are not considering whether, and to which degree, certain subsystems can be consider as separated from others and subject to completely autonomous dynamics.
  
The point we want to make is that normally we would not consider as natural or as acceptable a theory where the behavior of complex systems would help explain the behavior of simpler ones. As we remarked above, the fundamental theories of nature we are used to possess this clear directionality: fundamental laws, concerning basic or simple systems, allow the derivation of other more specific laws governing complex ones, and this is of course a very natural state of affairs when we consider the way in which we learn things (or, more precisely, the way in which we tend to order things in our minds). 
However, nature need not adapt to our learning and organizational practices. The natural world, after all, is just one single interconnected entity, which we arbitrarily chose to divide into portions or sectors in order to facilitate our characterization of its features.  Circularity, in the sense of A leading to, or explaining, B, B leading to, or explaining, C, and C leading to, or explaining, A, is something we would tend to reject in a fundamental theory of nature; nevertheless, there might be features of the functioning of the world that might require such circularity for its full comprehension. We cannot {\it a priori} reject such possibility and, as it occurs with any scientific proposal, it only can be judged by the success or failure of its predictions. 

The epistemological point we are making is simply that we should be open to the possibility that our methods for the development of a description of the world might need to accommodate such circularity. That is, we must consider the possibility that such circularity might be a required feature if we are ever going to attain a deeper understanding of the basic functioning of nature, (after all, the linearity of the explanatory schemes we normally use is related to the particular way our minds seem to work: i.e., by observing and learning from experiences whose records accumulate in our memories with the passage of time \cite{Callender1,Callender2,Callender3}). In fact some ideas containing features of this type have appeared in various kinds of physical proposals: the first one of course is the original bootstrap idea in particle physics, 
but a more recent proposal involving such circularity can be found in Lee Smolin's considerations regarding the mechanism fixing the value of the fundamental constants in our universe (see \cite{Smolin}). 
 
Here we want to consider the possibility that something of this sort might underlie the very nature of what we have so far called the fundamental collapse or wave function reduction process. As we have seen, those collapses represent the fundamental source of randomness in nature as well as the place where information is lost (and where new information is created). One might then inquire what are these collapse events, and the first answer one would consider is that they are fundamental aspects of nature, not to be understood in terms of anything else. However, when we consider the basic features of such events, it is hard not to think of them in analogy with black holes. In fact, if, as we have argued, black hole evaporation actually leads to information loss and breakdown of unitarity, that process would possess the essential features we have associated with the fundamental wave function collapse events.
 
Thus, what we want to contemplate is the following: according to quantum theory, looking at it through the path integral approach, all possible trajectories contribute to any process. Therefore, when dealing with ``quantum space-time,'' these trajectories will inevitably involve some arbitrarily small black holes (formed in connection to appropriately localized density fluctuations) that will rapidly evaporate. As a result, the evaporation process will lead to localized sources of randomness and information loss. That is, virtual black holes might also lead to information loss and breakdown of unitary evolution in essentially all evolution processes. Note that, as we mentioned above, Hawking noted that this virtual black holes effect would be expected if black hole evaporation involves such departures from standard quantum evolution \cite{Haw:00}. That is, information loss and breakdown of unitarity would be an aspect of physics that is present in all situations. Of course, it must occur at a level that does not conflict with the well established success of quantum physics in the many experimental tests. The idea we want to consider is that not only black hole information loss and breakdown of unitarity can be understood as the result of these fundamental collapse processes but that the collapse process itself can be regraded as some microscopic and virtual version of black holes formation and evaporation.
 
It shall be clear that when actually attempting to formulate such theory, the fundamental collapse events must somehow be postulated {\it ab initio} as part of the axioms of the theory, but one should also be guided by the requirement that, as one develops the theory and studies its consequences, a self-consistent picture should emerge where the fundamental dynamics is recovered in the appropriate limit. That is, the properties of the resulting dynamics for relatively large (emergent) black holes should be well described in terms of semiclassical general relativity and quantum field theory, and the early dynamics of such black holes, should conform to the standard analysis leading to Hawking's conclusions. However, when considering black holes on smaller scales, the resulting behavior should, according to these ideas, approach that of the postulated fundamental collapse events.

We should note that it is not completely clear whether, in the end, this kind of circularity will need to be incorporated. If, for instance, some path integral approach leads to the conclusion that black hole evaporation is associated with breakdown of unitarity, then there might be only a few plausible possibilities for the stochastic law governing the evolution of the quantum state, and these possibilities might correspond to collapse models.\footnote{A study of a range of possibilities for a stochastic fundamental law of evolution is presented in \cite{Bassi2}} Of course, such a path integral approach would need to be non-standard because one could argue on general grounds that any proposal based on ordinary quantum theory would have to yield a completely unitary evolution. At any rate, the answer as to whether or not the circularity aspects are in fact required will likely be found only when we have a complete and satisfactory theory incorporating the description of space-time and matter at a fundamental quantum level -- we present them here only as a possibility we find both promising and appealing.\footnote{We would like to thank an anonymous referee for leading us to explore this issue in some detail.}

\section{Summary}
\label{Sum}
We have discussed some of the major conceptual difficulties that arise in the interface of the quantum and gravitational realms, ranging from ones directly connected with quantum gravity, such as the problem of time, to others usually considered to lie in the relative beginning setting where a simple perturbative treatments ought to work (such  as the emergence of the seeds of structure in the inflationary universe). We have argued that a modification of quantum mechanics, along the lines of what are known as dynamical reduction theories, seems to contain the basic ingredients to deal with all such difficulties. Of course, we are not at this point claiming to have at hand a specific version of such theories which successfully accomplishes all what is required, and that can be taken as a real candidate for a theory of nature. We hope, however, that the present work will help motivate others to join the search for a theory of that kind, and to explore related problems that might either help in narrowing the search or to find ways to invalidate the generic proposal considered here. 
 
We have seen how a dynamical collapse theory can provide the mechanism by which an initial state that is homogeneous and isotropic evolves into one that is not, and thus can possess the seeds of cosmic structure despite the fact that the Hamiltonian controlling the unitary part of the evolution can only preserve the initial symmetry. We have also discussed how the incorporation of a stochastic aspect in the evolution of a diffeomorphism invariant theory like general relativity removes the basic ingredient that lies behind the so called ``problem of time'': the complete equivalence between the space-time information contained in any  space-like hypersurface.  Finally we have shown how a theory that incorporates at the fundamental level a loss of information and a lack of unitarity in its dynamics removes, at least at the qualitative level, one of the most serious problems afflicting our understanding the physics of black holes, namely, the information loss paradox. We should acknowledge the strong influence that Penrose's views and arguments have had on our ideas. In particular, we endorse his position regarding the likelihood that when attempting to put together quantum theory and gravitation, one will have to modify both (rather than just trying to adapt gravity to the standard quantum rules as is done in most approaches to quantum gravity).   


We should end up by noting that the de-Broglie-Bohm theory (dBB), which we also regard as a promising observer-independent version of quantum mechanics, might represent a strong competitor to our ideas if applied to the problems here considered, (note however that while collapse theories admit genuinely relativistic generalizations, dBB seems only to admit relativistic generalizations with (presumably undetectable) preferred frames). In fact, some work in these directions has been done already. For instance, applications of dBB to cosmology include \cite{dBB-Inflation1,dBB-Inflation2,dBB-Inflation3,dBB-Inflation4,dBB-Inflation5,dBB-Inflation6}, and, in particular, in \cite{dBB-Inflation1,dBB-Inflation2,dBB-Inflation3} it is argued that, at least in the cosmological setting, one might question the validity of the equilibrium prescription for the initial dBB distribution (of course not all works which apply dBB to cosmology share this non-equilibrium assumption). Such out-of-equilibrium proposal is based on cosmological considerations, like the fact that we do not have access to an ensemble of universes. As a result, these works conclude that dBB can yield some interesting deviations from the standard results, showing that the view that dBB must be seen just as a reinterpretation of standard quantum mechanics is not always fully warranted. 

Regarding the other issues we have touched upon, we note that in \cite{Gol.Teu:01} it is argued that the problem of time in quantum gravity might also be resolved within the context of dBB. As for the issue of black hole evaporation and loss of information, it is not unlikely that it might also be addressed within the scope of such theories, generating a plausible resolution. In fact, we believe it is quite likely that dynamical reduction theories and dBB are linked. That issue of course will need to be left for future works.
\begin{acknowledgements}
We would like to acknowledge partial financial support from DGAPA - UNAM projects IN107412 (DS), IA400312 (EO), and CONACyT project 101712 (DS).
\end{acknowledgements}

\bibliographystyle{spmpsci}       
\bibliography{biblioOCC.bib}
%
%

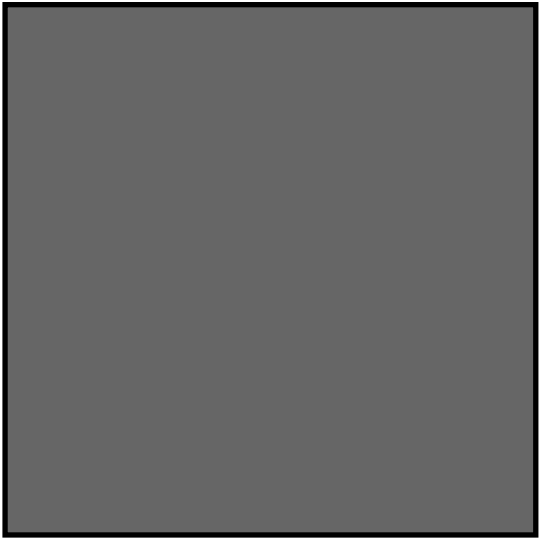
\end{document}